\documentclass[twocolumn,prl,amsmath,amssymb,showpacs,superscriptaddress]{revtex4-1}
\usepackage{epsf}      
\usepackage{graphicx}
\usepackage{color}
\usepackage{gensymb}
\usepackage{amsmath}
\usepackage{sidecap}

\begin{document}

\title{Structural and transport properties of La$_{1-x}$Sr$_x$Co$_{1-y}$Nb$_y$O$_3$ thin films}

\author{Rishabh Shukla}
\affiliation{Department of Physics, Indian Institute of Technology Delhi, Hauz Khas, New Delhi-110016, India}
\author{Ajay Kumar}
\affiliation{Department of Physics, Indian Institute of Technology Delhi, Hauz Khas, New Delhi-110016, India}
\author{Sandeep Dalal}
\affiliation{Solid State Physics Laboratory, DRDO, Lucknow Road, Timarpur, Delhi, 110054, India}
\author{Akhilesh Pandey}
\affiliation{Solid State Physics Laboratory, DRDO, Lucknow Road, Timarpur, Delhi, 110054, India}
\author{R. S. Dhaka}
\email{rsdhaka@physics.iitd.ac.in}
\affiliation{Department of Physics, Indian Institute of Technology Delhi, Hauz Khas, New Delhi-110016, India}

\date{\today}
\begin{abstract}

We present the structural and transport properties of La$_{1-x}$Sr$_x$Co$_{1-y}$Nb$_y$O$_3$ ($y=$ 0.1 and $x=$ 0; $y=$ 0.15 and $x=$ 0.3) thin films grown on (001) orientated single crystalline ceramic substrates to investigate the effect of lattice induced compressive and tensile strain. The high resolution x-ray diffraction measurements, including $\theta$-2$\theta$ scan, $\Phi$-scan, and reciprocal space mapping, affirm single phase; four-fold symmetry; good quality of deposited thin films. The atomic force micrographs confirm that these films have small root mean square roughness in the range of $\sim$0.5--7~nm. We observed additional Raman active modes in the films owing to the lowered crystal symmetry as compared to the bulk. More interestingly, the temperature dependent dc-resistivity measurements reveal that films become insulating due to induced lattice strain in comparison to bulk, however for the larger compressive strained films conductivity increase significantly owing to the higher degree of $p-d$ hybridization and reduction in bandwidth near the Fermi level.

\end{abstract}

\maketitle

\section{\noindent ~Introduction}

The aspiration to control material's properties with external perturbation motivates recent research on the strain driven discovery of exotic physical properties \cite{RameshNM07, BrinkmanNM07, YamadaScience04}. Remarkably, the strain effect in epitaxial thin films have been established as an alternative tool for mechanical and chemical pressure for tuning the material's properties  via governing their effective correlation (U/W) and electronic bandwidth (W) \cite{VoigtPRB03, VankoPRB06}. In complex oxides, the lattice mismatch induces octahedral distortions and rotations, which results in the origin of exceptional phenomena like ferromagnetism (FM), superconductivity, spin-state and metal-insulator transitions, etc. \cite{ChakhalianNP06, AhnNature04, LocquetNature98, CavigliaPRL12, VailionisAPL14, FuchsPRB07, DhakaPRB15}. In this family, perovskite cobaltite LaCoO$_3$ (LCO) tempted interest of researchers since many decades due to thermally driven spin-state transition anomalies of Co$^{3+}$ ions near 85 and 500~K \cite{KorotinPRB96,HaverkortPRL06}. The spin-state of Co$^{3+}$ evolves from a nonmagnetic low-spin (LS, t$_{2g}^6$e$_g^0$, S=0) to paramagnetic intermediate spin (IS, t$_{2g}^5$e$_g^1$, S=1) state near 85~K, and a metal to semiconducting transition around 500~K advocate that Co$^{3+}$ ions present in the mixed state of IS and high-spin (HS, t$_{2g}^4$e$_g^2$, S=2) state \cite{HeikesPhysica64}. The  energy of IS state is slightly higher than the LS state and much lower than HS state, due to the strong hybridization of Co 3$d$ and O 2$p$ orbitals. The crossover between spin-states with temperature emerge due to an exquisite interplay between crystal field splitting ($\Delta_{cf}$) and Hund's exchange energy (J$_{ex}$). As the $\Delta_{cf}$ is very sensitive to the Co--O bond length and Co--O--Co bond angle \cite{FreelandAPL08, VoigtPRB03, ZhouPRB05}, a subtle octahedral perturbation induced by biaxial strain can profoundly alter the spin states of Co ions, thereby affecting the magnetism and ferroic order of LCO films \cite{FuchsPRB07, HerkoltzPRB09, GuoSA19}. The partially occupied e$_g$ orbitals in IS state induce strong Jahn--Teller (JT) distortion and exhibit a long range e$_g$ orbital order, which lowers crystal symmetry of LCO from distorted rhombohedral (R-3c) to monoclinic (I2/a) \cite{KorotinPRB96, YamaguchiPRB97, MarisPRB03, LoucaPRL03}. This lowering in crystal symmetry induces new infrared and Raman active phonon modes, which were forbidden for the R-3c symmetry \cite{IshikawaPRL04, SeikhJMS04}. Recently, we have investigated the effect of cationic substitution(Sr and/or Nb) on structural, magnetic, and transport properties of bulk LCO \cite{ShuklaPRB18, ShuklaJPCC19}. The substitution of one Nb$^{5+}$(electron-doping) in LCO changes two Co$^{3+}$ into Co$^{2+}$ and stabilize them in HS state with a $^4$T$_1$ ground state \cite{ShuklaPRB18}. In addition to this, Nb$^{5+}$ ions drive system into insulating state, where a d$^0$ configuration of Nb$^{5+}$ ions and increased concentration of Co$^{2+}$ ions hinders carrier transport through the Co/Nb-O channel \cite{ShuklaPRB18, OygardenJSSC12}. Moreover, substitution of Sr$^{2+}$(hole-doping) transform Co$^{3+}$ into Co$^{4+}$ and stabilize ferromagnetic (FM) metallic ground state with a mixture of IS-HS state of Co$^{3+}$ ions and LS state of Co$^{4+}$ ions, respectively \cite{PrakshJALCOM18, SenarisJSSC95}. More interestingly, the co-substitution of Sr$^{2+}$ and Nb$^{5+}$ in a ratio of 2:1 preserve valence and spin-states of Co$^{3+}$ ions, a cluster spin-glass behavior emerges, and insulating nature prevails in the sample signifying the dominance of Nb$^{5+}$ ions in transport mechanism \cite{ShuklaJPCC19, AjayPRB20}.

Notably, the first principles calculations predicted that a half-metallic FM nature will evolve in the strained LCO thin films \cite{RondinelliPRB09, GuptaPRB09, PosadasAPL11, HsuPRB12}, whereas a FM insulating nature has been observed in experiments until now \cite{FreelandAPL08, LiAIP18}. The strain induced tetragonal distortion of CoO$_6$ octahedra in LCO owing to the change in Co-O-Co bond angle only (no change in bond-length) from 163$^{\rm o}$ toward 180$^{\rm o}$ leads to the stronger hybridization of the Co 3$d$ and O 2$p$ orbitals and hence results to the increased bandwidth and even overlap of the e$_g$-derived bands \cite{HsuPRB12, FuchsPRB07, FuchsPRB08, MehtaJAP09, MehtaPRB15, WangPRB19}. Intriguingly, strain dependent transport properties of LCO investigated by Li {\it et al.} reveal insulating nature of films and enhanced conductivity for the larger compressive strain \cite{LiAIP18}. This change in the conductivity of thin films with induced lattice strain is an effect of structural perturbation. Therefore, alteration in the size, shape and connectivity of the CoO$_6$ octahedra is crucial and responsible for the octahedral tilting and distortions, which leads to the redistribution of electrons in the t$_{2g}$ and e$_{g}$ states and strongly affect the structural, magnetic, and transport responses in the material. The tensile strain dependent spin-states observed by Yokoyama \textit{et al.} \cite{YokoyamaPRL18} and strain driven switchable orbital polarization to maximize the functionality of oxide heterostructures  \cite{GuoPRM19, PetrieAFM16} are the excellent findings in this direction. The strain induced octahedral modifications in LCO films lead to its practical applications in photocatalytic activity \cite{LiuJALCOM19}, gas sensing applications \cite{LiuASS18}, and establish it as a potential cathode material in the solid oxide fuel cells \cite{MiyoshiSSI16}.

Therefore, in this paper we study the effect of lattice induced strain on the structural, vibrational, and dc-transport properties of La$_{1-x}$Sr$_x$Co$_{1-y}$Nb$_y$O$_3$ epitaxial thin films, grown on the (001) oriented single crystalline substrates. High resolution x-ray diffraction measurements; $\theta-2\theta$, $\phi$-scan, reciprocal space mapping; were performed to confirm that the films are of single phase and highly oriented good quality. The Raman spectroscopy data reveal new phonon modes in thin films due lowering in the dimensionality. The dc-resistivity of all films exhibits insulating/semiconducting nature and they follow Arrhenius type conduction in the higher temperature region while 3D variable range hopping (VRH) model in the lower temperature regime.

\section{\noindent ~Experimental}

\begin{figure}
\includegraphics[width=3.5in]{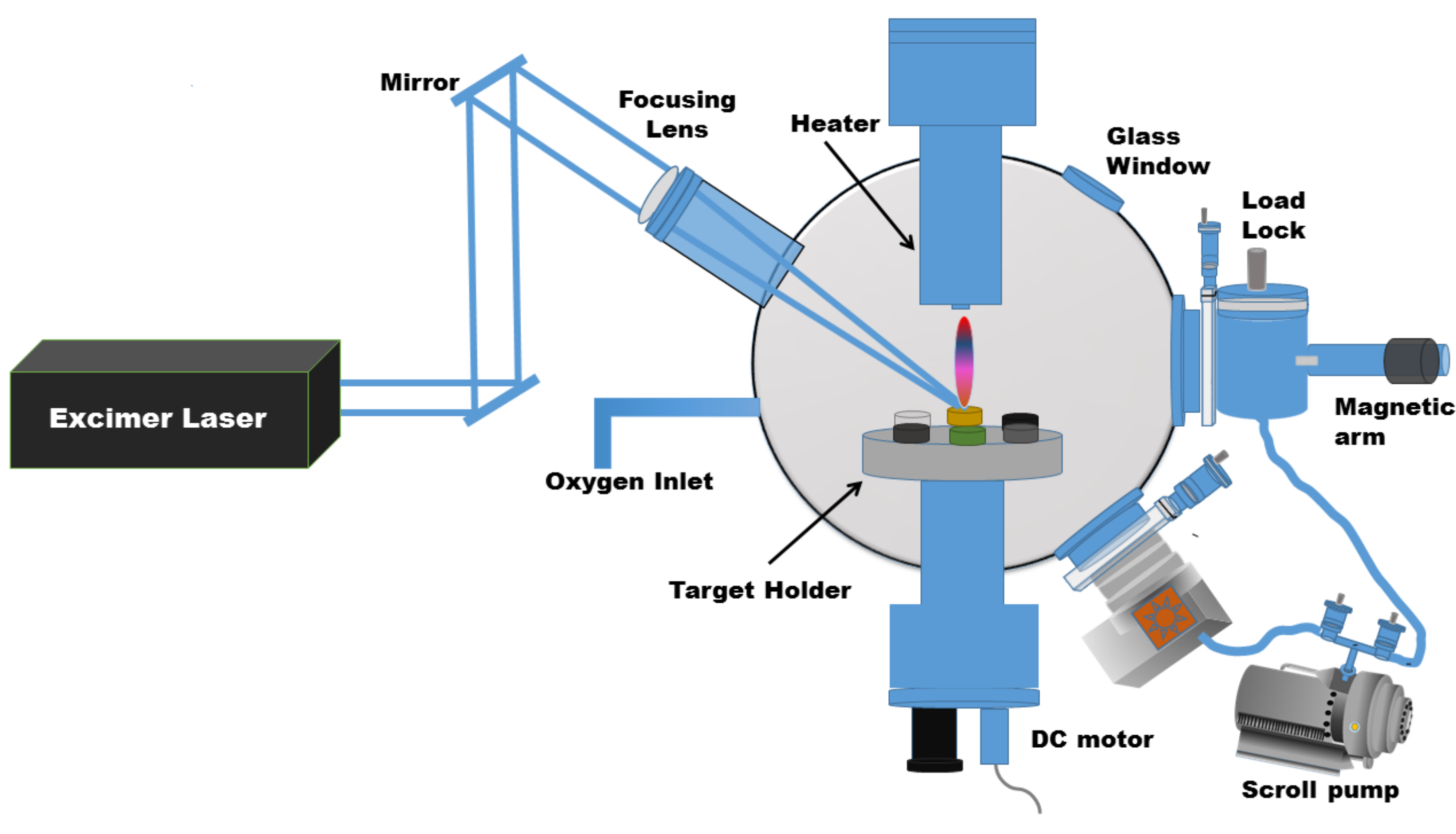}
\caption[]{A schematic diagram of the ultra high vacuum compatible pulsed laser deposition setup equipped with an excimer laser of wavelength 248~nm.}
\label{fig:PLD}
\end{figure}

In order to grow heterostructures of complex oxide and metallic systems, we have set-up a new pulsed laser deposition (PLD) system, see a schematic diagram in the Fig.~\ref{fig:PLD},  equipped with an ultraviolet excimer laser from Coherent GmbH, Germany. The base pressure of the order of 10$^{-10}$~mbar was achieved in the main growth chamber, which is equipped with air-cooled turbo molecular (nEXT400D) and scroll (nXDS15i) pumps from Edwards vacuum, UK. The ultra high vacuum (UHV) ensures for the growth of clean and high quality thin films of complex oxides and inter-metallic compounds. The target flange is used to mount six bulk targets on a circular stage in the sixfold geometry where each target holder has a circular hole to dissipate heat during the longer depositions. In order to ablate the target material, we use a special geometry to guide the laser beam with the help of mirrors and lenses considering the diverging nature of the beam. The laser spot focused on the target is of rectangular shape with the dimensions of 1.5$\times$4~mm$^2$ and have all sharp edges for the homogeneous ablation of the material. In addition, each target holder has two motions: one is the rotation on its own axis (controlled through a dc-motor) and other is raster motion in the defined angle range (controlled through the step motor). These combined motions of target further help to ablate the target uniformly to avoid the formation of craters, which in turn avoid the possibility of particulates during the deposition process. The substrate flange equipped with a resistive plate heater, which is used to achieve upto 850$^{\rm o}$C temperature on the substrate as per the requirement for high quality thin film growth. The distance between target and substrate is about 5.5~cm. In addition, we have a small load-lock chamber attached with a manual gate valve to the main growth chamber. We use a magnetic transfer arm for insertion/extraction of the substrate without interrupting vacuum in the growth chamber. We have optimized all the parameters including photon flux, repetition rate, distance between target and substrate as well as oxygen partial pressure for the growth of high quality thin films studied in this paper.

Using our new PLD system, we have grown epitaxial thin films of La$_{0.7}$Sr$_{0.3}$Co$_{0.85}$Nb$_{0.15}$O$_3$ (LSCNO), and LaCo$_{0.9}$Nb$_{0.1}$O$_3$ (LCNO) on the (001) oriented single crystalline substrates of LaAlO$_3$ (LAO), (LaAlO$_3$)$_{0.3}$(Sr$_2$TaAlO$_6$)$_{0.7}$ (LSAT) and SrTiO$_3$ (STO) having dimensions around 3$\times$5$\times$0.5~mm$^3$. We have synthesized ceramic targets via solid-state reaction method, more details of preparation and characterization can be found elsewhere \cite{ShuklaJPCC19,ShuklaPRB18}. A KrF excimer laser of wavelength 248~nm is used for the deposition at repetition rate of 3~Hz with energy density of 1--2~J/cm$^2$. We kept the substrate temperature of about 700$^{\rm o}$C and 0.25~mbar oxygen partial pressure during the deposition. After the deposition, films were annealed at the deposition temperature for 15~mins maintaining the oxygen partial pressure around 250~mbar. Then, cooling to room temperature was done with a rate of 5~K/min to ensure the oxygen stoichiometry. For the determination of structural parameters, high resolution x-ray diffraction data were collected using PANalytical X'pert Pro MRD HR-XRD with Cu-K$\alpha$ radiation ($\lambda=$1.5406~\AA) using two-circle diffractometer in Bragg-Brantano geometry. Also, a four-circle high resolution x-ray diffractometer was used for in-depth analysis of structure, orientation and epitaxy of films. Atomic force microscopy images were recorded with a Bruker iconXR system in tapping mode. A Horiba confocal microscope was used to measure Raman spectra with an excitation wavelength of 514.5~nm and a laser power of 0.5~mW. Also, dc--transport measurements in a four-probe configuration were performed with a constant current of 0.1~$\mu$A using a physical property measurement system (PPMS) from Quantum Design, USA.

\section{\noindent ~Results and Discussion}

\begin{figure}
\includegraphics[width=3.5in]{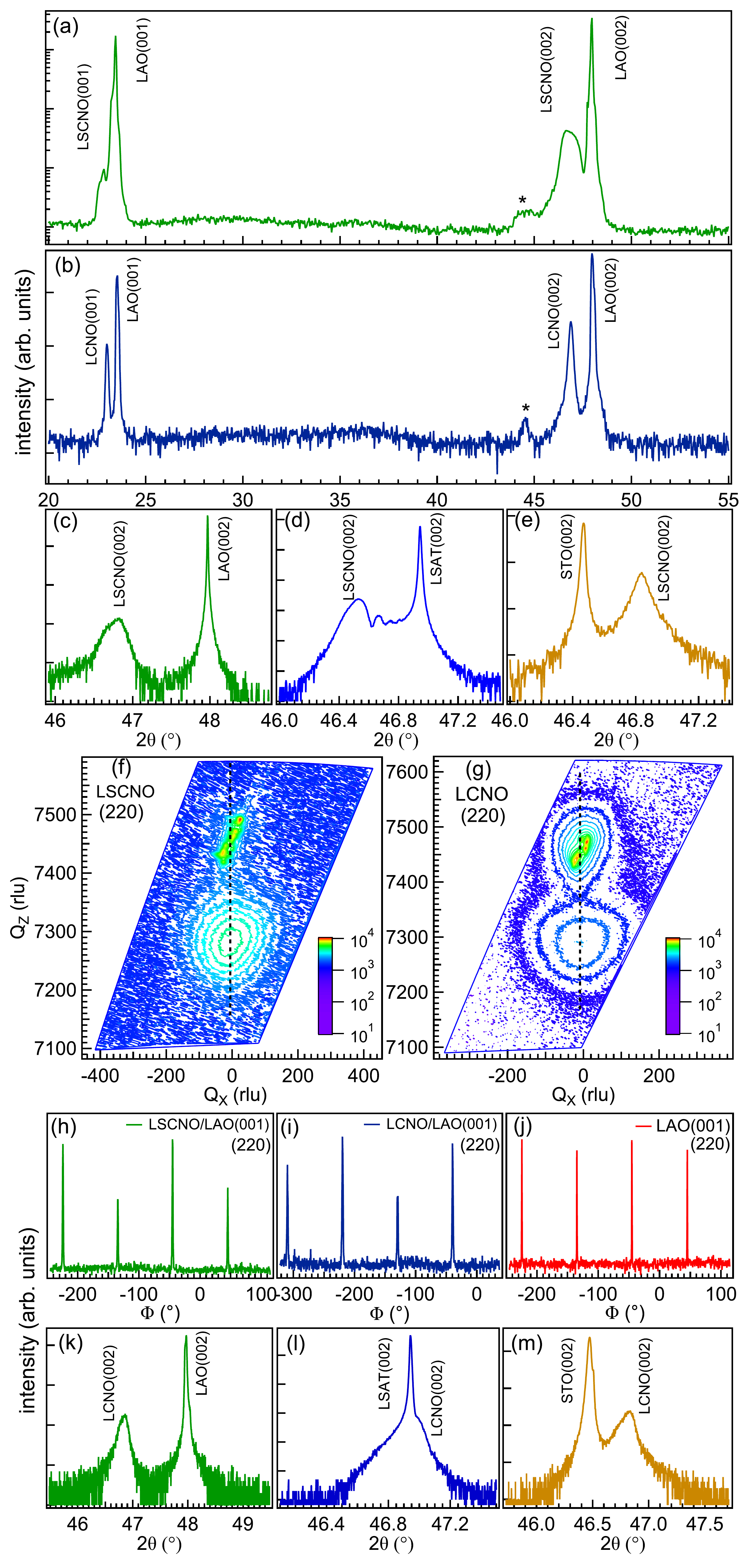}
\caption{High resolution room temperature x-ray diffraction: (a, b) out of plane $\theta$-2$\theta$ patterns of LSCNO and LCNO thin films on LAO(001) substrate, respectively, (c--e) the slow scans for the LSCNO films near the (002) reflection for the LAO, LSAT and STO substrates recorded in the triple axis mode, (f, g) reciprocal space mapping of the LSCNO and LCNO thin films on the LAO substrate along the asymmetric (220) peak, (h--j) $\phi$-scans of the LSCNO and LCNO films on LAO along with the bare LAO substrate about an asymmetric (220) plane, (k--m) the slow scans for the LCNO films near the (002) reflection for the LAO, LSAT and STO substrates recorded in the triple axis mode, respectively.}
\label{fig:XRD}
\end{figure}

In order to characterize the structural properties and quality of thin films we performed high resolution x-ray diffraction (XRD) measurements at room temperature in various geometries. In Figs.~\ref{fig:XRD}(a, b), we show the full range out-of-plane XRD patterns recorded in the Bragg-Brantano geometry ($\theta-2\theta$ mode) for LSCNO and LCNO thin films on LAO(001) substrate, respectively. It is clear that only (\textit{00l}) reflections are present, which confirms that films are oriented along the growth direction of substrate and no impurity/unwanted phase is observed except a small peak (marked by asterisk) from the sample holder. Further, in the slow scans around (002) reflection, as shown in Figs.~\ref{fig:XRD}(c--e), we observed peak from LSCNO films at lower 2$\theta$ value than the LAO and LSAT substrates indicate the compressive strain, whereas in case of STO substrate the film peak is on the higher 2$\theta$ value manifests tensile strain in the films. The in-plane compression (elongation) of the lattice parameters give rise to the increase (decrease) in the c-axis parameter, which reflects in the peak at the lower (higher) 2$\theta$ values. The film peaks have large full width at half maximum (FWHM) as compare to the substrate peaks because of the smaller thickness of the films (120~nm). Note that due to the lattice mismatch between the substrate and film, the lattice parameters of the films are altered in comparison to their bulk counterparts. The in-plane and out-of-plane pseudo-cubic lattice parameters of bulk LSCNO sample are 3.876~\AA (a$_{pc}$=b$_{pc}$) and 3.839~\AA (c$_{pc}$) \cite{ShuklaJPCC19}. On the other hand, we found the out-of-plane parameter to be 3.881, 3.901, and 3.876~\AA~for the films deposited on LAO, LSAT and STO substrates, respectively. The out-of-plane lattice strain using the formula $\epsilon_{zz}$=[($c_{film}$-$c_{bulk}$)/$c_{bulk}$]$\times$100\% [where c$_{film}$ is calculated from the slow scans of XRD data, c$_{bulk}$ is the pseudo-cubic lattice parameter of the sample in bulk form] and expected misfit [(a$_{bulk}$-a$_{subs}$)/$a_{subs}$, where a$_{bulk}$ is the pseudo-cubic lattice parameter of the bulk sample and a$_{subs}$ is the lattice parameter of the substrates] between the substrate and bulk sample are calculated and presented in the Table~\ref{tab:Strain}. 
\begin{table}[h]
  \centering
  \caption{Lattice mismatch induced strain and fitted parameters from the temperature dependent resistance data of LSCNO films grown on different single crystal substrates.}
 \vskip 0.2 cm 
  \label{tab:Strain}
   \begin{tabular}{|c|c|c|c|c|c|}
  	\hline
   Substrate& a$_{sub}$&Misfit&Strain& 3D-VRH & Arrhenius\\
   &(\AA)&(\%)&$\epsilon_{zz}$(\%)& N(E)(eV$^{-1}$cm$^{-3}$)& E$_a$(meV)\\
   \hline
    LAO&3.790 &-2.27 &+1.1&3.1$\times$10$^{20}$& 183$\pm$1.5\\
    \hline
    LSAT& 3.868&-0.21 &+1.62 &1.3$\times$10$^{20}$  & 207$\pm$1.5\\
    \hline
    STO& 3.905&+0.74&+0.96&1.5$\times$10$^{20}$ & 198$\pm$1\\
    \hline
  \end{tabular}
\end{table}
Here the out-of-plane axis stretches for the compressive strained LSCNO/LAO and LSCNO/LSAT films, which results in the peak at lower 2$\theta$ value as compare to the c--parameter of the bulk. However for the tensile strained film on STO substrate we did not see contraction in the c--parameter, this inconsistent behavior is possibly due to the strain relaxation or twinning in the film. Interestingly, we observe the superlattice reflections known as Kiessig/Laue fringes in case of LSCNO/LSAT in the vicinity of the Bragg peak [see Fig.~\ref{fig:XRD}(d)], which confirm that these films are of high quality. The film thickness (D) can be calculated using the following formula \cite{KrostSpringer96} 
\begin{equation}
D = \frac{(L_m-L_n)\lambda}{2(sin\theta_m-sin\theta_n)}
\end{equation}
where $L_m$ and $L_n$ are the order of superlattice reflections and $\theta_m$ and $\theta_n$ are the angles of these diffraction peaks, respectively. Using the separation of the Kiessig fringes observed in the (002) reflection of LSCNO/LSAT in Fig.~\ref{fig:XRD}(d), we estimate the film thickness of about 120~nm, which is consistent with the one measured using stylus profilometer (not shown). 

Also, we have performed reciprocal space mapping along an asymmetric reflection (220) orientation and plotted in the (Q$_X$, Q$_Z$) plane, where Q$_X$ and Q$_Z$ are the components of the scattering vector \textbf{Q} (Q=4$\pi$Sin$\theta$/$\lambda$) in parallel to film plane and perpendicular to it, defined as the, Q$_X$ = 1/$\lambda$[cos($\omega$)-cos(2$\theta-\omega$)] and Q$_Z$ = 1/$\lambda$[sin($\omega$)+sin(2$\theta-\omega$)], respectively. Here, $\omega$ is angle between the incident x-ray and crystal surface, $\theta$ is known as an angle between incident x-ray and scattering lattice planes of crystal, and $\lambda$ is the wavelength of incident x-ray. The reciprocal space maps are shown in Figs.~\ref{fig:XRD}(f, g). More than one peaks observed in the RSM contour for the substrate indicates the off-axis orientation, which is well known as the twinned structure of the LAO crystal \cite{WangJCW97}. Further, we have performed the $\phi$-scan along the asymmetric (220) plane for the deposited thin films LSCNO, LCNO and substrate LAO, which are presented in Figs.~\ref{fig:XRD}(h-j), respectively. The four peaks in each pattern manifest four fold symmetry with 90$\degree$ intervals, suggesting for the single crystalline films, with well defined in-plane orientation. The two different sets of lattice planes for the LSCNO/LCNO films and LAO substrate occur at the same azimuthal angle $\phi$, which indicates that the unit cell of the films is well aligned in the basal planes of the LAO substrate in a cube-on-cube fashion, see Figs.~\ref{fig:XRD}(h, j). In case of the LCNO film, its bulk counterpart consists of rhombohedral and orthorhombic phases in a ratio of $\sim$80:20, respectively \cite{ShuklaPRB18}, whereas the deposited LCNO films are phase pure and epitaxial, which demonstrates the advantage and novelty of pulsed laser deposition. The slow scans near the (002) reflection of the LCNO films deposited on the LAO, LSAT and STO substrate are shown in the Figs.~\ref{fig:XRD} (k-m), respectively. The LCNO/LAO shows a compressive strain; whereas LCNO films on LSAT and STO are tensile strained, having out of plane parameters 3.876, 3.854 and 3.878~\AA, respectively. In Figs.~\ref{fig:XRD}(l), the peak from LCNO film overlap with the LSAT substrate peak, which is due to their similar lattice parameters.

\begin{figure}[h]
\includegraphics[width=3.4in]{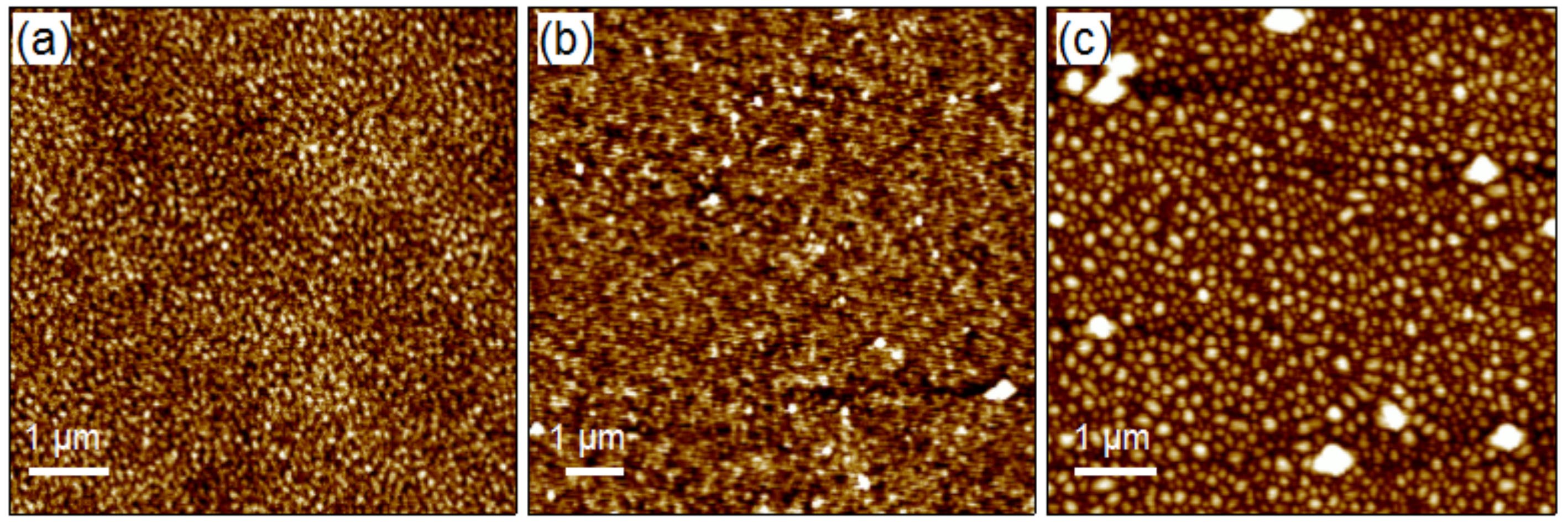}
\includegraphics[width=3.4in]{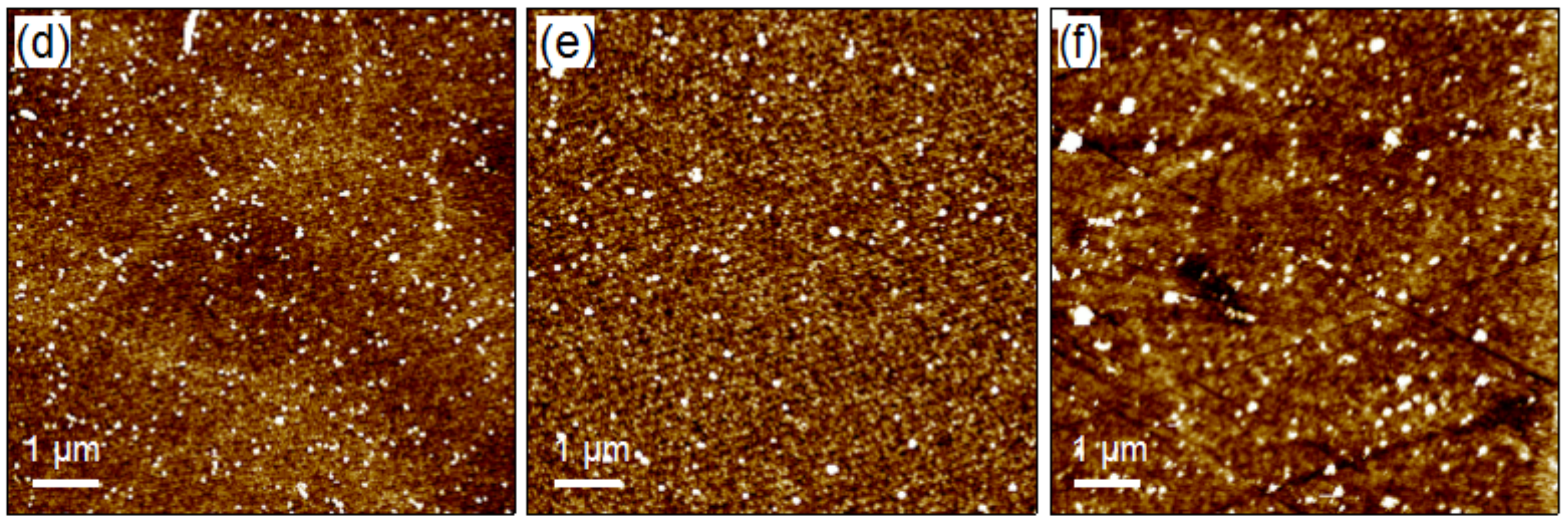}
\caption{Atomic force microscopy images of the surface of the LSCNO and LCNO films on the (a, d) LAO, (b, e) LSAT, and (c, f) STO substrates, respectively.}
\label{fig:AFM}
\end{figure}

To check the surface topography of these deposited films, we use atomic force microscopy (AFM) to record images of the LSCNO and LCNO films deposited on the substrates LAO, LSAT and STO, as shown in Figs.~\ref{fig:AFM}(a-f), respectively. These images indicate that the films on LAO and LSAT are of high quality in layer by layer growth mode, whereas the films on STO show relatively larger roughness. The estimated root mean square roughness values are 0.5, 3, and 7~nm for LSCNO films and 3.9, 2.8, and 4.8~nm for LCNO films on the LAO, LSAT, and STO substrates, respectively, which confirm the good quality of the films.

\begin{figure}[h]
\includegraphics[width=3.2in]{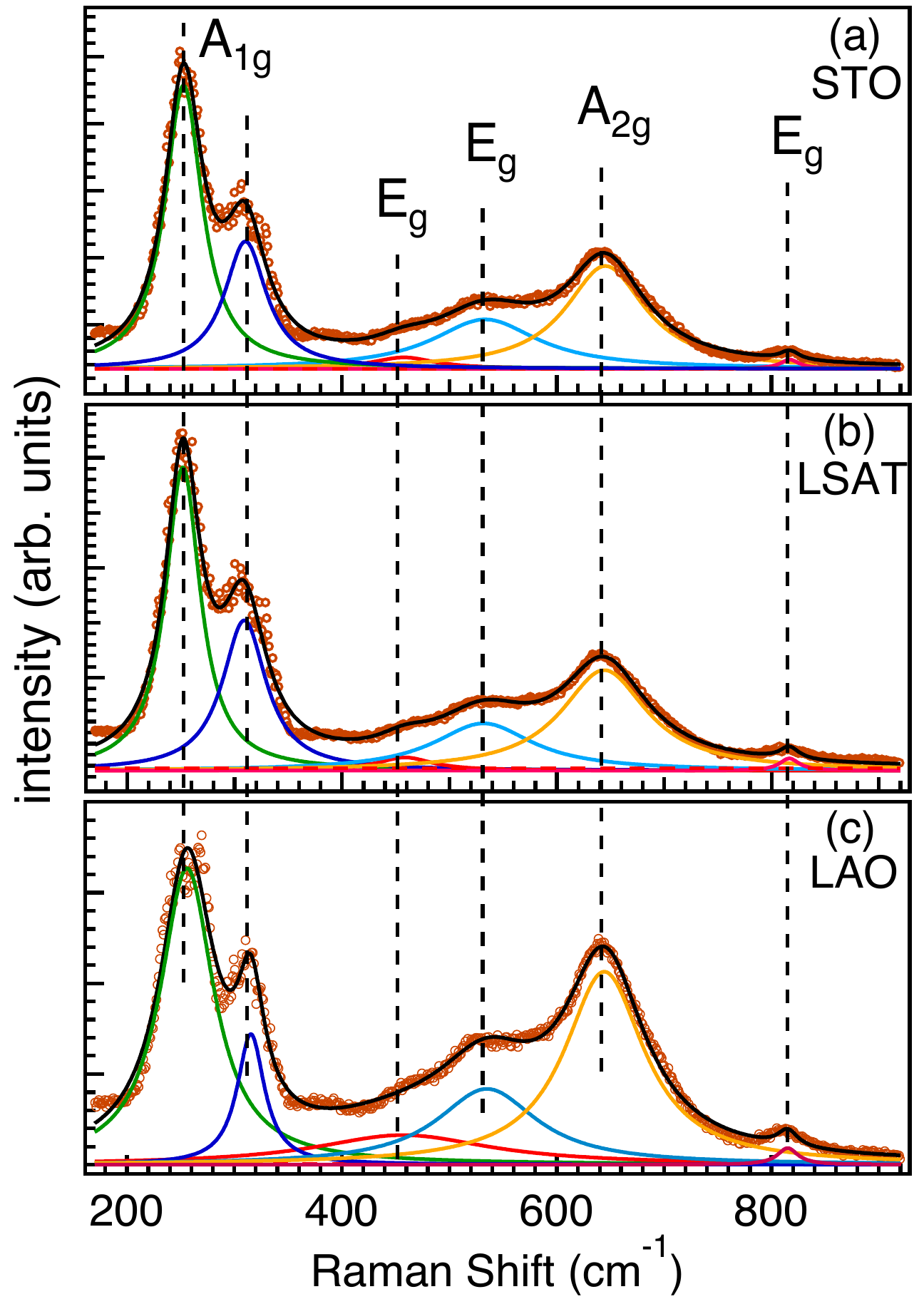}
\caption{Lorentzain peak shape fitted room temperature Raman spectra of the deposited LSCNO thin films on (a) STO, (b) LSAT, and (c) LAO substrates using an excitation wavelength of 514.5~nm.}
\label{fig:Raman1}
\end{figure}

\begin{figure}
\includegraphics[width=3.3in]{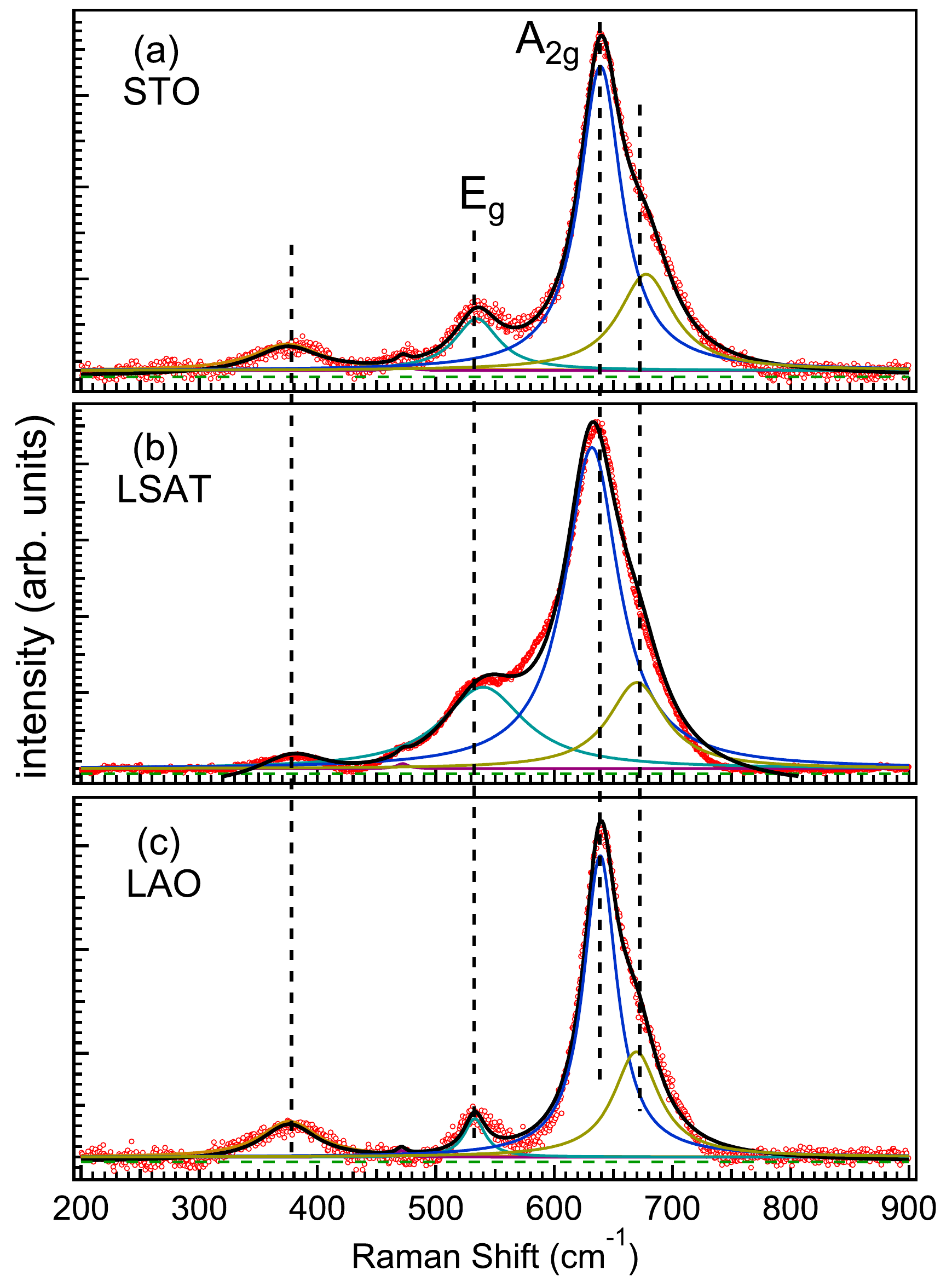}
\caption{De-convoluted room temperature Raman spectra of LCNO thin films on (a) STO, (b) LSAT, and (c) LAO substrates using an excitation wavelength of 514.5~nm are fitted with a Lorentzian function.}
\label{fig:Raman2}
\end{figure}

\begin{figure*}
\includegraphics[width=7.1in]{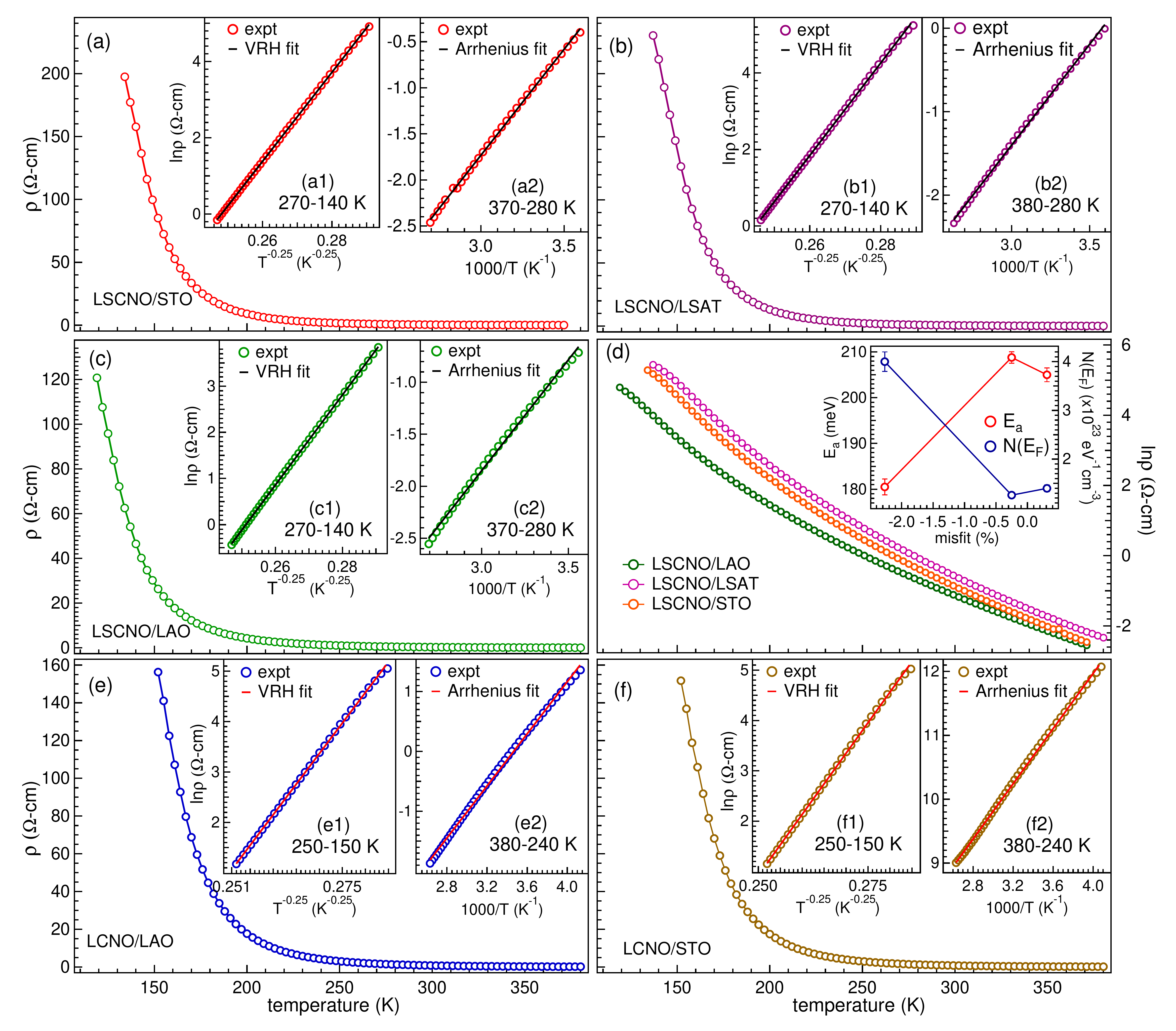}
\caption{Temperature dependent resistivity of LSCNO thin films grown on (a) LAO, (b) LSAT, and (c) STO substrates, (d) logarithmic comparison of resistivity for all LSCNO films on different substrates, and resistivity of LCNO thin films on (e) LAO, and (f) STO substrate. The insets in (a-c, e, f) exhibit fitting with the Arrhenius model in high temperature region and 3D variable range hopping (VRH) in low temperature region, and inset in (d) shows the strain dependence of activation energy (E$_a$) and density of states near the Fermi level N(E$_{\rm F}$) of films.}
\label{fig:RT_LSCNO}
\end{figure*}

In order to understand the effect of strain on the vibrational properties of thin films, the room temperature Raman spectra are shown for the LSCNO thin films in Fig.~\ref{fig:Raman1} along with the fitted components using Lorentzian peak shape. We observe two strong Raman modes A$_{1g}$ (corresponds to the rotations of CoO$_6$ octahedra around the hexagonal [001] direction) and A$_{2g}$ at 230 and 650~cm$^{-1}$ wave numbers along with weaker E$_g$ modes, as marked in the Fig.~\ref{fig:Raman1}. We note that the parent compound LaCoO$_3$ crystallizes in rhombohedral structure with the R$\bar3$c space group (point group D$^6_{3d}$, Z=2) in ambient conditions and can be seen in reference to a simple cubic perovskite (space group Pm$\bar3$m) structure by rotating the adjacent CoO$_6$ octahedra in opposite direction around the body diagonal direction [111]. This structure exhibits the octahedral rotations, that is classified as a$^-$a$^-$a$^-$ in the Glazer notation \cite{GlazerAC72}, which manifest that the adjacent octahedra in the lattice are out of phase tilted in equal magnitude, i.e., tilting with same angle but opposite in the direction. Here, the La atoms occupy 2a(1/4,1/4,1/4) Wyckoff positions and participate in the four phonon modes (E$_g$+E$_u$+A$_{2u}$+A$_{2g}$), Co atoms occupy 2b(0,0,0) Wyckoff positions and also participate in the four modes (2E$_u$+A$_{2u}$+A$_{1u}$), and oxygen atoms occupy 6e($x$,$x$+1/2,1/4) site, where $x$ is the free parameter that sets BO$_6$ rotation angle and participate in the twelve phonon modes ($A_{1g}+2A_{2g}+3E_g+A_{1u}+2A_{2u}+3E_u$) \cite{IshikawaPRL04, AbrashevPRB99, GnezdilovLTP03}. In these only five modes (A$_{1g}$ and 4E$_g$) are Raman active, whereas 3A$_{2u}$+5E$_u$ are the infrared active and acoustic modes. There exist 5 silent/inactive vibrational modes 3A$_{2g}$+2A$_{1u}$. However, it has been reported that for the Co$^{3+}$ ions to be in IS state, Jahn-Teller effect dominates and results in the lowering crystal symmetry to monoclinic $I2/a$, which give origin to the zone-folding into the k=0 point and roughly double the number of phonon modes \cite{IshikawaPRL04}. 

In the present case, the A$_{1g}$ and E$_{g}$ Raman peaks are related to the collective modes of the oxygen octahedral network. A weak Raman mode around 460~cm$^{-1}$ correspond to the pure E$_g$ bending mode of oxygen, where alternate oxygen ions in the octahedra at equivalent Wyckoff position stretches in the opposite direction. The peak $\approx$530~cm$^{-1}$ is related to the E$_g$ quadruple stretching mode, which indicates that the oxygen atoms at the corner of CoO$_6$ octahedra stretches in the opposite directions in the pair, i.e., two pairs of oxygen ions asymmetrically stretches with equal magnitude in the vertically succeeding octahedra \cite{IshikawaPRL04, GouPRB11}. The Raman mode at $\approx$640~cm$^{-1}$ correspond to the A$_{2g}$ breathing mode in which six Co-O bond vibrates in the outward and inward direction. This A$_{2g}$ mode is inactive/symmetry forbidden for the rhombohedral symmetry; however, this mode become active for the monoclinic symmetry $I2/a$ and have highest scattering intensity owing to the strong electron-phonon interactions \cite{IshikawaPRL04,HongJPPCL13}. The presence of Co$^{3+}$ ions in the IS state and e$_g$ orbital order owing to the partially occupied orbitals is accredited to induce the strong J-T distortion in the system and hence lowering the local crystal symmetry to monoclinic $I2/a$ \cite{MarisPRB03}. It is consistent as in bulk LSCNO the Co$^{3+}$ stabilizes in both IS and HS states with equal ratio \cite{ShuklaJPCC19}. For the LSCNO/LAO film sample, we observe an increase in the intensity of A$_{2g}$ mode as compared to the LSCNO/STO film. The change in the Raman modes with lattice induced strain are notably from the reduction in the Co--O bond strength and change in the Co--O--Co bond angle, as reported by Fuchs \textit{et al.} \cite{FuchsPRB08}. Moreover, in Fig.~\ref{fig:Raman2} we present the Raman spectra for the LCNO films, which show similar modes as observed in bulk counterpart \cite{ShuklaPRB18}. However, a new Raman mode near 380~cm$^{-1}$ is appeared, and the E$_g$ and A$_{2g}$ modes shifted towards lower wavenumber due to strain induced stretching of the Co--O bond in the thin film samples as compared to the bulk. It is also interesting to note that the A$_{1g}$ modes around 300~cm$^{-1}$ are present only in LSCNO films (Fig.~\ref{fig:Raman1}) and not in LCNO (Fig.~\ref{fig:Raman2}). This may signifies the role of Co valence state as there is only Co$^{3+}$ present in LSCNO \cite{ShuklaJPCC19}; whereas, LCNO has both Co$^{3+}$ and Co$^{2+}$ valence states \cite{ShuklaPRB18}. 

Further, to elucidate the effect of lattice induced biaxial strain on the electronic transport properties we have recorded the temperature dependent resistivity data ($\rho$ vs T) and presented in Figs.~\ref{fig:RT_LSCNO}(a--c) for LSCNO and Figs.~\ref{fig:RT_LSCNO}(e, f) for the LCNO films. We found a semiconducting/insulating behavior for all thin films owing to the negative temperature coefficient of the resistivity similar to their bulk counterpart \cite{ShuklaJPCC19, OygardenJSSC12}. We have compared the $\rho$ vs T data of LSCNO thin films on logarithmic scale in Fig.~\ref{fig:RT_LSCNO}(d), which clearly demonstrate that LSCNO/LAO has the lower resistivity in comparison to the other two films, which is related to strong $p$--$d$ hybridization (strong overlapping of the metal-ligand wavefunctions, i.e., Co 3$d$/Nb 4$d$ and O 2$p$ orbitals) \cite{FreelandAPL08, LiAIP18}. We were able to collect the resistivity up to the $\sim$100~K  because of higher value of resistance and compliance limit of the instrument. The resistivity behavior suggest that the transport mechanism in high temperature region is governed by the Arrhenius model (conduction by simple activation of charge carriers through the band gap between conduction and valence band). This conduction of charge carriers following the Arrhenius model can be estimated using following equation:
\begin{equation}
\rho(T) = \rho_0 exp(E_a/k_BT)
\end{equation}
where, $\rho_0$ and E$_a$ are the pre-exponential factor and activation energy, respectively. This activation energy is calculated from the linear fit of ln($\rho$) versus 1/T curve, as shown in the right insets in the resistivity versus temperature graphs in Figs.~\ref{fig:RT_LSCNO}.

However, we note here that the system deviates from the Arrhenius model below $\approx$260~K and indicate the existence of other conduction mechanism in the lower temperatures. We found that the low temperature regime conduction mechanism is dominated by the Mott's variable range hopping (VRH) model, where conduction of carriers is followed by the mediation of oxygen anions via Co/Nb--O path \cite{ShuklaPRB18}. The transport behavior following 3D--VRH model can be described by the below equation:
\begin{equation}
\rho(T)=\rho_0 exp{(T_0/T)^{1/4}}
\end{equation}
where, T$_0$ and $\rho_0$ are the characteristic temperature and constant factor, respectively. The value of T$_0$ can be defined with the formula $T_0= 18/k_BN(E_{\rm F})\alpha^3$, where N(E$_{\rm F}$) depicts the density of states near the Fermi level and $\alpha$ is the localization length for hopping of charge carriers \cite{ShuklaPRB18}. This value of $\alpha$ can be approximated to the average Co/Nb--O bond length of material, which is 1.955~\AA~ for LSCNO and 1.936~\AA~ for LCNO. Here, first we calculate the value of T$_0$ from linear fit of ln($\rho$) versus T$^{-0.25}$ curve, as shown in the left insets of Figs.~\ref{fig:RT_LSCNO}, and the obtained value of T$_0$ is used to estimate the N(E$_{\rm F}$). The estimated activation energy for the charge carriers and density of state near the Fermi level for LSCNO films are presented in Table-\ref{tab:Strain}. We found that for the highly compressive LSCNO/LAO film activation energy barrier of the charge carriers decrease and density of states near the Fermi level increases in comparison to the other films [see inset of Fig.~\ref{fig:RT_LSCNO}(d)] and bulk (204$\pm$2~meV, 2$\times$10$^{20}$~eV$^{-1}$cm$^{-3}$) \cite{ShuklaJPCC19}. Thus, the conduction of the charge carriers for LSCNO/LAO film become easier, owing to the strong metal ligand overlap between Co 3$d$/Nb 4$d$ and O 2$p$ orbitals. These observations suggest that at higher strain in LSCNO films, we are able to distort its (Co/Nb)O$_6$ octahedra via changing the unit cell parameters of the layer and hence the dynamics of charge carriers. Similarly, the calculated values of activation energy for LCNO/LAO and LCNO/STO films are 192.5$\pm$1 and 189.5$\pm$1~eV, respectively; likewise values for density of states near the Fermi level are close to each other and found to be 1.35$\times$10$^{20}$ ~cm$^{-3}$eV$^{-1}$.

The mean value of hopping energy (E$_{h}$) and mean hopping distance (R$_h$) can be calculated using the following relations \cite{ChandraSSC15,PaulPRL83}:
\begin{eqnarray}
R_h (T) &=& [9\alpha/8\pi kTN(E_F)]^{1/4}~cm\\
E_h (T) &=& [3/4\pi R_h^3N(E_F)]~eV
\end{eqnarray}
The estimated values of R$_h$(T) and E$_h$(T) for the LSCNO films are in the range of 19--23~nm and 110--140~meV, respectively,  while for LCNO films these values are $\approx$17~nm and $\approx$370~meV, respectively. These calculated parameters satisfy the requirement for the Mott--VRH conduction mechanism, which are $\alpha^{-1}R_h\gg$1 and E$_h\gg k_B$T \cite{PaulPRL83}. Moreover, reduction in the value of R$_h$ exhibit decrease in the bandwidth near Fermi level and results in the enhancement of hopping of charge carriers \cite{McNallyNPJQM19}. We found that the lowest value of R$_h$ for LSCNO/LAO and that supports the reason for the increase in the conductivity of the sample.

\section{\noindent ~Conclusions}

We have successfully deposited LSCNO and LCNO epitaxial thin films on the (001) oriented LAO, LSAT and STO substrates having different strain. Our high resolution x-ray diffraction data in $\theta-2\theta$ mode confirm that the films are single phase, $\phi$-scan measurements exhibit four-fold symmetry in coherence to the substrates and exhibit the cube-on-cube growth. In addition, the reciprocal space mapping measurements present the detailed picture and reveal that films are of good quality. The root mean square roughness obtained from AFM manifest that the films have smaller roughness. The vibrational properties reveal new Raman active modes due to lowering of local crystal symmetry in the films as compare to the bulk counterparts. The temperature dependent dc-transport measurements indicate that the films exhibit semiconducting/insulating nature with Arrhenius type conduction and 3D variable range hopping model in the higher and lower temperature regions, respectively. Interestingly, we found that higher compressive strain favors for conducting owing to the enhanced metal-ligand overlap and decrease in the bandwidth near Fermi level.





\section*{\noindent ~Acknowledgments}

This work was financially supported by SERB-DST through Early Career Research (ECR) Award with project reference no. ECR/2015/000159 and planning unit of IIT Delhi through Seed Grant with reference no. BPHY2368. RS and AK gratefully acknowledge the DST-Inspire and UGC, India for fellowship, respectively. We thank Dr. Mahesh Chandra for useful discussions and Dr. Subhash Pai for his help in designing our PLD system. We acknowledge the physics department of IIT Delhi for PPMS EVERCOOL-II and nano research facilities for providing the AFM, stylus profilometer, and Raman spectrometer. We use a high temperature Nabertherm furnace for target preparation, which is supported by BRNS through DAE Young Scientist Research Award project sanction no. 34/20/12/2015/BRNS.

\end{document}